\documentclass[structabstract]{aa}  
\usepackage{graphicx}
\usepackage{txfonts}
\usepackage[stable]{footmisc}
\usepackage{natbib}

\newcommand {\ca} {Ca\,{\small II}}
\newcommand {\hi} {{\rm H}\,{\small\rm I}}
\newcommand {\kms} {\,{\rm km\,s}^{-1}}

\newcommand {\kpc} {\,{\rm kpc}}

\newcommand {\mo}{\,{\rm M}_\odot}
\newcommand {\lo}{\,{\rm L}_{\odot}}

\newcommand {\gsim}{\,\lower.7ex\hbox{$\;\stackrel{\textstyle>}{\sim}\;$}}
\newcommand {\lsim}{\,\lower.7ex\hbox{$\;\stackrel{\textstyle<}{\sim}\;$}}

\newcommand {\fe}{[{\rm Fe/H}]}

\begin{document}

\title{Life at the Periphery of the Local Group: the kinematics of the 
Tucana Dwarf Galaxy\footnotemark[1]{}
}

\author{Filippo Fraternali
  \inst{1}
  \and
Eline Tolstoy
  \inst{2}
  \and
Mike Irwin
  \inst{3}
\and
Andrew Cole
  \inst{4}
}
\institute{
 Department of Astronomy, University of Bologna, 
  via Ranzani 1, 40127, Bologna, Italy\\
  \email{filippo.fraternali@unibo.it}
    \and
Kapteyn Astronomical Institute, University of Groningen, Postbus 800, 
9700 AV Groningen, Netherlands.\\
  \email{etolstoy@astro.rug.nl} 
  \and
Institute of Astronomy, University of Cambridge, Madingley Road, 
Cambridge CB3 0HA, UK.\\
  \email{mike@cam.ast.ac.uk}
 \and
 School of Mathematics \& Physics
University of Tasmania, Private Bag 37 Hobart, 
Tasmania 7001 Australia.\\
\email{andrew.cole@utas.edu.au}
}

  \date{Received ; accepted }

  % \abstract{}{}{}{}{} 
  % 5 {} token are mandatory

  \abstract
  % aims heading (mandatory)
  {}
   {
Dwarf spheroidal galaxies in the Local Group are usually located close
to the Milky Way or M31. Currently, there are two clear
exceptions to this rule,  and the Tucana dwarf galaxy is the most
distant at almost 1 Mpc from the Milky Way. Our aim is to learn more
about the nature of Tucana by measuring its radial velocity and
internal kinematics.
  }
  % methods heading (mandatory)
 {
Using the VLT/FORS2 spectrograph in multi-object mode we were able to 
measure the velocities of 23 individual Red Giant Branch stars in and 
around Tucana using the Ca triplet absorption lines. 
From this sample, 17 reliable members have been identified. 
 }
 % Results heading (mandatory)
 {
We measured the systemic velocity and dispersion of Tucana to be 
$v_{\rm hel} = +194.0\pm4.3\kms$ and $\sigma_{\rm l.o.s.}=15.8^{+4.1}_{-3.1}$
respectively.
These measures are derived after removing the signature of rotation 
using a linear gradient of 
$6.5 \times R/R_{\rm core}\pm2.9\kms$, which corresponds to a rotation 
of $\approx16\kms$ at the reliable limit of our data.
Our systemic velocity corresponds to a receding velocity from the
barycentre of the Local Group of $v_{\rm LG} = +73.3\kms$.  
We also determined the mean metallicity of Tucana to be 
\fe$= -1.95\pm0.15$ with a dispersion of $0.32\pm0.06$~dex.
}
  % conclusions
  {
Our study firmly excludes any obvious association of Tucana with the
\hi\ emission in the vicinity and shows that Tucana is a genuine dwarf
spheroidal, with low metallicity stars, no gaseous ISM and no recent
star formation.  The present location and relatively high recession
velocity are consistent with Tucana having been an isolated Local
Group galaxy for the majority of its existence.  
%The relatively high
%central velocity dispersion we measure together with existing surface
%brightness photometry suggest a high mass-to-light ratio,
%$M/L\approx100$ and a mass for Tucana of $\approx6\times 10^7\mo$ 
%within its nominal tidal radius.
  }
 

\maketitle

\section{Introduction} \label{s:intro}

\footnotetext[1]{Based on FORS1 and FORS2 observations collected at
the European Southern Observatory, proposals 63.N-0560 and 69.B-0305}

Due to their faintness, dwarf spheroidal galaxies are, at the moment,
observed and studied in detail only in the Local Group.  They are
characterized by complex past star formation histories, the lack of
present-day gas and thus a lack of ongoing star formation
\citep[e.g.,][]{mat98}.  Most of them tend to lie very close to the
giant spirals [the Milky Way and M\,31] and interactions with these
could be the reason for the lack of gas. There are two currently known
exceptions: Cetus, located at 775~kpc from the Milky Way and 680~kpc
from M~31 \citep{whi99, lew07} and, Tucana, which lies some 900~kpc
from the Milky Way and 1350~kpc from M~31 \citep{lav92,sav96}.  A
third galaxy, And~XVIII may fall in the same category
\citep{mccon08}.

The association of Tucana with the Local Group is relatively recent
\citep{lav92}.  Color-magnitude diagrams obtained with CCD photometry
\citep{sav96,cas96} show no evidence for recent star formation and
suggest that Tucana has been completely dormant for the last
$8-10$~Gyr. There is also clear evidence for an ancient stellar
population, as demonstrated by an extended Horizontal Branch
\citep{et00,hol06} and RR~Lyrae stars \citep{bern08}.  The average
metallicity, derived by comparing the color-magnitude diagram (CMD)
with globular clusters, has been estimated to be between $\fe=-1.6$
and $-1.8$ \citep{sav96,cas96}.  These physical properties suggest
Tucana is a typical dwarf spheroidal 
(dSph) galaxy but with the peculiarity of being located at
a large distance from the luminous members of the Local Group, and
$\approx$1175 kpc from the Local Group barycentre, see Table
\ref{t:tucana}.

%Figure 1
\begin{figure*}[!ht]
\begin{center}
\includegraphics[width=0.85\textwidth]{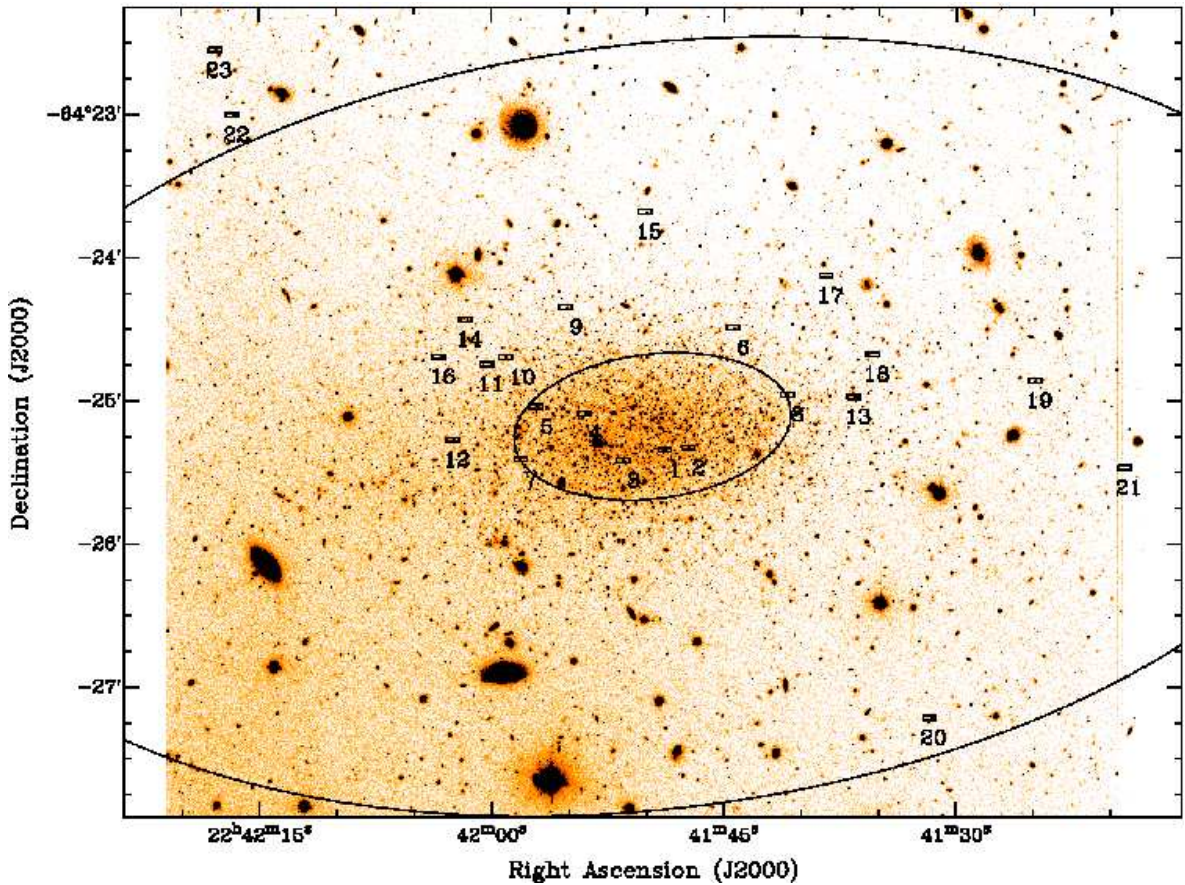}
\caption{
Optical image (R-band, FORS1) of Tucana overlaid with the positions of
the slits covering the stars for which we obtained spectra with FORS2.
The small and large ellipses show the locations and orientations of
the geometric core and ``tidal'' radii. North is up and East to the left.
\label{f:stars}}
\end{center}
\end{figure*}

A search for neutral gas (\hi\ at 21-cm) associated with Tucana has
been carried out by \citet{oos96} using the Australian Telescope
Compact Array (ATCA).  They found an upper limit on the \hi\ mass
associated with Tucana of $M_{\rm HI}/L_{\rm B}$ of $\approx$ 10$^{-2}$
which is consistent with values for other dSphs \citep{mat98,
young2000}.  In the same data, an \hi\ cloud was found at a projected
distance of 15$'$ from Tucana (3.4~kpc at its distance) with mean
heliocentric velocity of 130$\kms$.  This velocity is close to that
of the Magellanic Stream cloud complexes in that region
\citep{putman03} and \citet{oos96} suggested this as the most probable
association of the cloud.  On the other hand, other authors have
argued in favour of an association with Tucana \citep[e.g.][]{blitz00,
bouch06}.

In this paper we present results from FORS2 multi-object spectroscopy
of a sample of Red Giant Branch (RGB) stars in and around Tucana (see
Fig.\ \ref{f:stars}).  Using these data we determine an accurate
optical radial velocity of the galaxy (Section~\ref{s:systemic}) by
fitting \ca\ triplet absorption lines.  We also study the internal
kinematics of Tucana (Section~\ref{s:kinematics}) and estimate the
metallicity of the RGB stars using \ca\ triplet equivalent widths
(Section~\ref{s:metallicity}).  We then discuss the possible
association of Tucana with the \hi\ cloud and its unusual location
within the Local Group (Section~\ref{s:discussion}).

\begin{table}[!ht]
\caption{{Tucana Dwarf physical parameters}
\label{t:tucana}}
\centering    
\begin{tabular}{lcc}
\hline\hline
Parameter & Tucana Dwarf & Ref.\\
\hline
Type & dSph & - \\
{RA, DEC} (J2000) & 22 41 49.6, $-$64 25 11 & (1) \\
{l, b} ($^{\circ}$) & 322.9, -47.4 & - \\
{Distance} (kpc) & 880 & (2) \\
{L$_{\rm V}$} ($10^5 \lo$) & 5.5 & (3) \\
v$_{\rm sys}$ & 194.0 & (4) \\
{R$_{\rm core}$} ($''$) & 42 & (3) \\
{R$_{\rm tidal}$} ($''$) & 222 & (3) \\
{p.a.} ($^{\circ}$) & 97 & (3) \\
{e} & 0.48 & (3) \\
V$_{HB}$ & 25.4 & (4)\\
Conversion 1$''$/pc & 4.3 & - \\ 
\hline                  
\end{tabular}\\
(1) \citet{lav92};
(2) \citet{cas96, sav96}; 
(3) \citet{mat98}; 
(4) this work.\\
\end{table}

\section{Observations and data reduction}
\label{s:observations}

The observations of Tucana were obtained in service mode using the
UT4/FORS2 spectrograph at the VLT over 3 nights in August 2002 in the
multi-object (MXU) mode (see Table \ref{t:obs}).  We used the
GRIS\_1028z grism and the OG590 filter in order to cover the spectral
region of the \ca\ triplet absorption lines.  A total of 45 slits were
placed on individual stars over a spatial region of about 8$'$
($\approx$2~kpc) centred on Tucana.  Out of the 45 slits, 30 were located
on positions of stars selected to lie on the RGB in the CMD (see
Figure~2), and the remaining targets were randomly selected to fill
otherwise empty slits.  The spatial size of each slit is
1.2$''\times8''$, the dispersion scale is 0.86~\AA\,pixel$^{-1}$
($\approx$30$\kms$) leading to a velocity resolution (from the
dispersion of Gaussian-like unresolved sky lines) of about 50$\kms$.

%Figure 2
\begin{figure*}[!ht]
\begin{center}
\includegraphics[width=0.9\textwidth]{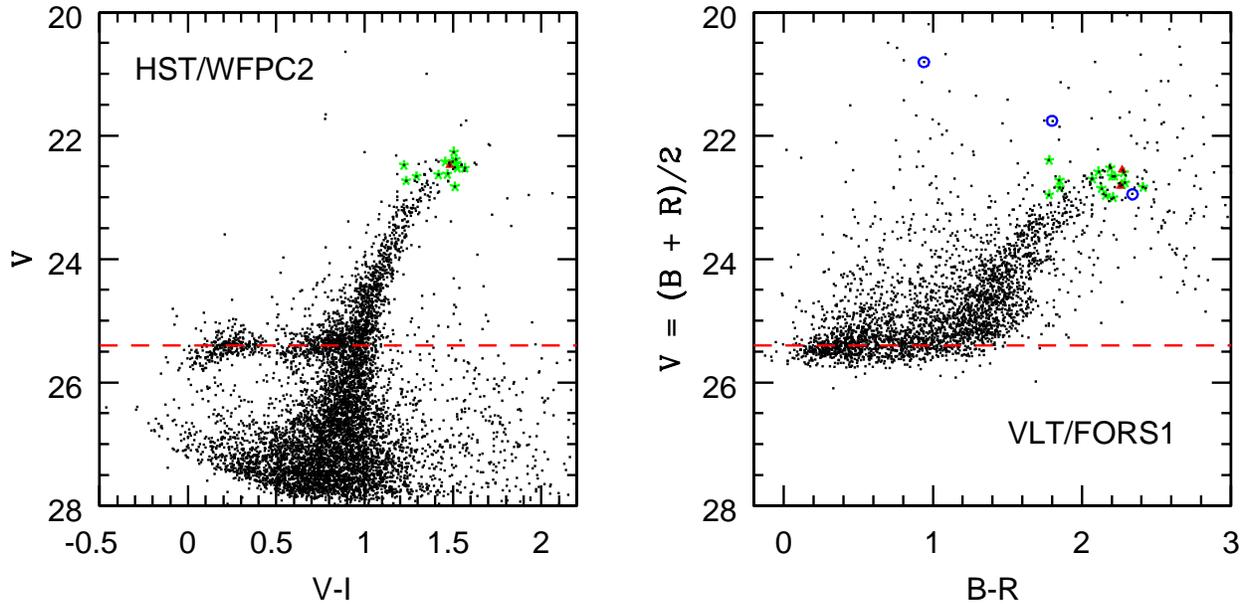}
\caption{
Colour-magnitude diagrams of Tucana, the one on the left comes from
archival HST/WFPC2 data \citep{hol06} and the one on the right comes
from UT1/FORS1 imaging \citep{et00}.  In both CMDs the targets
selected for spectroscopy are plotted as symbols different from the
underlying CMD. In both cases likely velocity members are plotted as
green star shaped symbols and non-members are plotted as blue open
circles. The stars which are probably members but which we do not
include in our analysis are plotted as red solid triangles.  Also
plotted on the CMDs is the position of V$_{HB} =25.4$, determined
visually from the HST CMD, and this is the value used in the
determination of the Ca II triplet metallicities.  Not all the stars
have both HST and FORS photometry, see Table~3, which is why not all
spectroscopic stars are plotted on both CMDs, one star does not appear
on either CMD (\#21).
\label{f:cmd}}
\end{center}
\end{figure*}

\begin{table}[!ht]
\caption{{Observational Log}
\label{t:obs}}
\centering
\begin{tabular}{ccccc}
\hline\hline
Date & UT (end) & Airmass & Seeing & Dataset ID \\ 
& (h:m:s) &  & ($''$) & (FORS2.2002-08-)\\
\hline

12/08/2002 & 04:09:42 & 1.383 & 0.98 & 12T04:09:42.883 \\
           & 04:36:32 & 1.347 & 0.88 & 12T04:36:32.163 \\
           & 05:11:55 & 1.316 & 0.71 & 12T05:11:55.313 \\
           & 05:38:44 & 1.303 & 0.66 & 12T05:38:44.081 \\

13/08/2002 & 03:14:54 & 1.483 & 1.15 & 13T03:14:54.375 \\
           & 03:41:41 & 1.425 & 1.03 & 13T03:41:41.247 \\

14/08/2002 & 01:37:57 & 1.822 & 1.03 & 14T01:37:57.100 \\
           & 02:04:44 & 1.699 & -    & 14T02:04:44.521 \\
           & 02:32:36 & 1.594 & -    & 14T02:32:36.985 \\
           & 02:59:27 & 1.513 & 0.72 & 14T02:59:27.785 \\
           & 03:30:20 & 1.439 & 0.77 & 14T03:30:20.610 \\
           & 03:57:08 & 1.39  & 0.85 & 14T03:57:08.873 \\

\hline                  
\end{tabular}
For each observation the exposure time is 1560 s.
\end{table}

Standard data reduction steps were carried out with IRAF (Image
Reduction and Analysis Facility) and with our own IDL routines.
Observations from the same night were combined together in order to
remove cosmic rays. In total 12 exposures of 1560~sec were made for a
total integration time of 5.2~hrs.  We first corrected for distortions
of the focal plane by fitting a 3rd order polynomial function to the
position of the objects.  Since the objects are much fainter than the
sky lines, we performed the wavelength calibration using the sky lines
themselves.  In order to determine the wavelength solutions, we fitted
polynomial functions of an order that depends on the number of
identified lines, but was typically of order 3 with the maximum a 4th
order polynomial.  The accuracy of the wavelength solution was tested
by fitting Gaussian functions to sky lines of well known wavelengths
around the \ca\ triplet region (8450-8700~\AA\,), after the wavelength
calibration had been applied.  The resulting errors are typically
$\approx5\kms$.  Looking at the through-slit images taken before and
after each spectroscopic sequence all the objects with successfully
extracted spectra were well centred in the slit and we did not need to
make additional corrections to the radial velocities for slit
mis-centering \citep[e.g.,][]{tol00}.

%Figure 3
\begin{figure}[ht]
\centering
\includegraphics[width=\columnwidth]{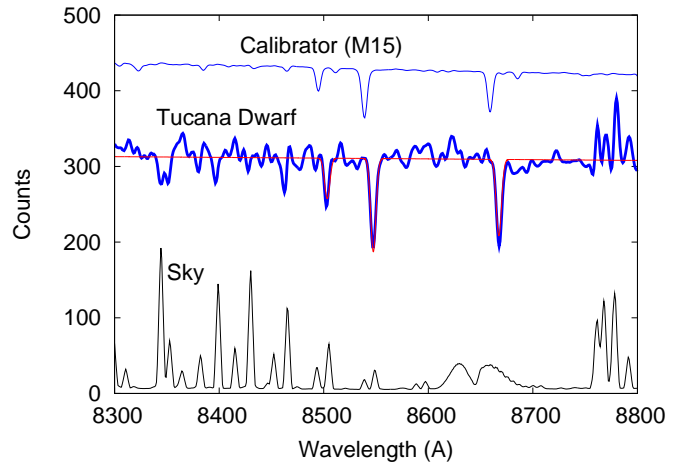}
\caption{
A sample spectrum (thick blue line) of a star in Tucana (\#4 in Table
\ref{t:stars}).  The thin (red) line overlaid shows our fit of the
\ca\ triplet absorption lines using Gaussian profiles.  The upper
spectrum (thin blue line) belongs to a star in the calibration
globular cluster M~15.  The bottom spectrum shows the sky emission
spectrum scaled down by a factor 30.  Note the clear shift between the
location of the \ca\ triplet in M~15 and Tucana.
\label{f:spectrum}}
\end{figure}

\begin{table*}[!ht]
\caption{{Physical parameters of stars in our FORS2 field}
\label{t:stars}}
\begin{center}
\begin{tabular}{lcccccccccccc}
\hline\hline
N. & R.A. & Dec & V & V$-$I & R & B$-$R & r$^a$ & S/N & $V_{hel}$ & Member? & Eq.\ Width$^b$ & [Fe/H]\\
 & (J2000) & (J2000) & (1) & (1) & (2) & (2) & (arcsec) & (\AA\,$^{-1}$) & ($\kms$) & & (\AA\,) & dex \\
\hline

1  & 22 41 48.8 & -64 25 20.4 & 22.48 & 1.22 & 21.80 & 1.85 & 10.8  & 16.1 & 172.0 $\pm$ 9.4 &Y&  4.11 $\pm$ 0.28  &  -1.82 \\
2  & 22 41 47.3 & -64 25 19.5 & 22.53 & 1.57 & 21.62 & 2.29 & 17.1  & 23.2 & 184.6 $\pm$ 5.7 &Y&  4.61 $\pm$ 0.20  &  -1.59 \\
3  & 22 41 51.5 & -64 25 25.0 & 22.64 & 1.42 & 21.89 & 2.16 & 18.7  & 22.1 & 200.7 $\pm$ 6.4 &Y&  3.66 $\pm$ 0.22  &  -1.98 \\
4  & 22 41 54.0 & -64 25 05.1 & 22.39 & 1.52 & 21.56 & 2.20 & 29.1  & 30.5 & 180.0 $\pm$ 5.5 &Y&  3.46 $\pm$ 0.14  &  -2.14 \\
5  & 22 41 57.1 & -64 25 02.4 & 22.82 & 1.51 & 21.79 & 2.13 & 49.3  & 19.5 & 187.5 $\pm$ 6.7 &Y&  3.84 $\pm$ 0.23  &  -1.85 \\
6  & 22 41 44.4 & -64 24 29.2 & 22.42 & 1.46 & -     & -    & 53.7  & 18.8 & 159.6 $\pm$ 10.0&Y&  2.94 $\pm$ 0.36  &  -2.36 \\
7  & 22 41 58.1 & -64 25 24.3 & 22.66 & 1.30 & 21.92 & 1.85 & 56.6  & 15.6 & 225.2 $\pm$ 6.9 &Y&  4.16 $\pm$ 0.29  &  -1.75 \\
8  & 22 41 40.9 & -64 24 57.2 & 22.62 & 1.47 & 21.90 & 2.21 & 58.0  & 14.5 & 205.7 $\pm$ 12.5&Y&  4.99 $\pm$ 0.41  &  -1.40 \\
9  & 22 41 55.2 & -64 24 20.6 & 22.42 & 1.50 & 21.53 & 2.11 & 62.1  & 28.6 & 204.8 $\pm$ 8.4 &Y&  3.43 $\pm$ 0.15  &  -2.14 \\
10$^c$ & 22 41 59.1&-64 24 41.6&22.48 & 1.48 & 21.69 & 2.26 & 68.2  & 25.1 & 234.5 $\pm$ 9.6 &Y?& 3.03 $\pm$ 0.21  &  -2.30 \\
11 & 22 42 00.3 & -64 24 44.5 & 22.73 & 1.24 & 22.07 & 1.78 & 74.3  & 14.2 & 192.7 $\pm$ 12.0&Y&  4.22 $\pm$ 0.32  &  -1.71 \\
12 & 22 42 02.5 & -64 25 16.2 & 22.26 & 1.51 & 21.43 & 2.19 & 83.7  & 18.8 & 232.8 $\pm$ 6.3 &Y&  3.96 $\pm$ 0.33  &  -1.95 \\
13 & 22 41 36.6 & -64 24 58.2 & 22.34 & 1.60 & 21.46 & 2.28 & 85.1  & 25.3 & 169.0 $\pm$ 7.0 &Y&  2.87 $\pm$ 0.18  &  -2.41 \\
14 & 22 42 01.7 & -64 24 25.8 & -     & -    & 21.55 & 2.22 & 89.2  & 25.6 & 195.1 $\pm$ 9.7 &Y&  2.80 $\pm$ 0.21  &  -2.39 \\
15 & 22 41 50.1 & -64 23 40.7 & -     & -    & 21.51 & 1.78 & 90.3  & 29.4 & 172.9 $\pm$ 7.3 &Y&  4.08 $\pm$ 0.14  &  -1.86 \\
16 & 22 42 03.4 & -64 24 41.3 & -     & -    & 21.63 & 2.41 & 94.2  & 22.2 & 200.2 $\pm$ 6.8 &Y&  5.38 $\pm$ 0.21  &  -1.17 \\
17 & 22 41 38.4 & -64 24 07.4 & 22.52 & 1.53 & 21.67 & 2.07 & 96.5  & 23.1 & 181.3 $\pm$ 9.2 &Y&  3.18 $\pm$ 0.22  &  -2.22 \\
18 & 22 41 35.4 & -64 24 40.4 & 22.45 & 1.52 & 21.66 & 2.27 & 96.9  & 21.1 & 197.5 $\pm$ 7.3 &Y&  4.40 $\pm$ 0.31  &  -1.70 \\
19 & 22 41 24.9 & -64 24 51.3 & -     & -    & 21.78 & 2.34 & 161.1 & 16.7 &  70.4 $\pm$ 8.1 &N&  3.51 $\pm$ 0.32  &  -     \\
20 & 22 41 31.7 & -64 27 12.6 & -     & -    & 20.34 & 0.94 & 168.0 & 36.3 &  78.1 $\pm$ 37.8&N&  1.11 $\pm$ 0.16  &  -     \\
21 & 22 41 19.1 & -64 25 27.8 & -     & -    & -     & -    & 200.2 & 18.1 & 234.9 $\pm$ 6.3 &Y?& 3.88 $\pm$ 0.37  &  -1.92$^d$ \\
22 & 22 42 16.7 & -64 22 59.9 & -     & -    & 21.42 & 2.27 & 219.2 & 26.8 & 201.6 $\pm$ 11.5&Y?& 3.53 $\pm$ 0.20  &  -2.06 \\
23 & 22 42 17.8 & -64 22 32.8 & -     & -    & 20.86 & 1.80 & 241.7 & 19.0 &   5.6 $\pm$ 5.6 &N&  4.19 $\pm$ 0.29  &  -     \\
\hline
\end{tabular}
\end{center}   
(1) From WFPC2 photometry, \citep{hol06}. (2) From FORS1 photometry,
\citep{et00}. $^a$ Projected distance from the centre of Tucana.  $^b$
Sum of the two strongest \ca\ lines. $^c$ Double star. $^d$ Using the
fiducial value of V=22.5, coming from the signal-to-noise (S/N) of the star.
\end{table*}

Sky subtraction was carried out by fitting a 1st or 2nd order
polynomial function to the object-free regions along the spatial
direction after correcting for the bending of the lines along the
spatial axis.  In order to improve the extraction, the 2D spectra were
then smoothed to a velocity resolution of about 60$\kms$ dispersion.
After sky subtraction, the spectra of the objects were extracted along
the spatial axis.  Gaussian functions were fitted to the data (after
binning along the dispersion axis) and the object spectra extracted by
summing a spatial region typically within $\pm$2$\sigma$ from the
central positions.  Although the slits have been straightened at this
stage, there may be a residual displacement (typically $\pm$1~pixel)
of the objects with respect to their average positions therefore the
objects have been further traced using a polynomial function.

Once the spectra have been extracted, the \ca\ triplet absorption
lines were used to determine the radial velocities of the stars by
fitting a triple-Gaussian function template with fixed relative
distances between the three Gaussians.  The relative intensity ratios
of the \ca\ lines were also fixed by calibrating them on the globular
cluster M~15.  Finally, we assumed an equal velocity dispersion for the
three lines.  With these assumptions the free parameters of the \ca\
fitting are: 1) the position along the dispersion axis (radial
velocity); 2) the global intensity scaling and the 3) dispersion of
the lines.  Two more parameters are used to fit the underlying
continuum, which is assumed to be linear in the region of interest.

\section{Results} \label{s:results}

Table \ref{t:stars} shows the stars for which we obtained an
acceptable fit for the \ca\ triplet lines 
(signal-to-noise per \AA\,~$> 14$).  The
spatial locations of these stars are shown in Fig.\ \ref{f:stars},
where N refers to the slit number which is also plotted in Fig.\
\ref{f:stars}.  The quality of these fits was also checked with the
cross-correlation of the \ca\ triplet-Gaussian model with the full
spectrum.  All the stars included in Table \ref{t:stars} show a clear
peak in the cross-correlation.  As an example of the quality of the
data, Figure \ref{f:spectrum} shows the spectrum obtained for star
\#4 in Table \ref{t:stars}.

\subsection{Systemic velocity and velocity distribution}
\label{s:systemic}

%Figure 4
\begin{figure}[ht]
\begin{center}
\includegraphics[width=6cm, angle=-90]{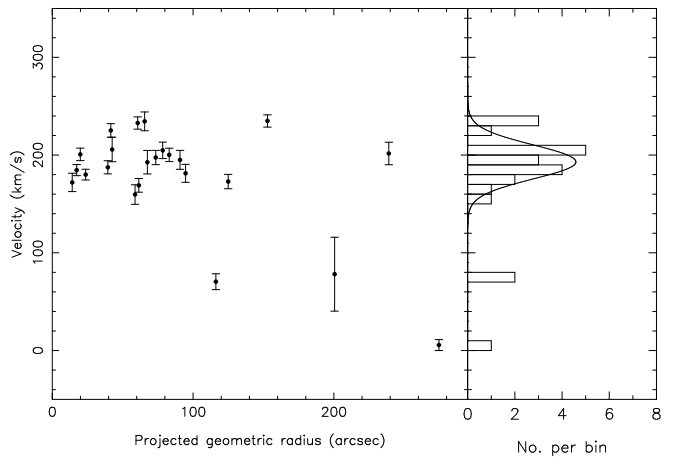}
\caption{
Left panel: the projected geometric elliptical radial distribution of the
velocities for all the stars in Table \ref{t:stars}.  Note the clear
divisions between central likely members, outer possible members and
lower velocity foreground stars.  Right panel: a histogram of the
velocity distribution of the most probable 17 members sample, with the
maximum likelihood Gaussian model fit, $v_{\rm hel} = +193.0\pm4.9\kms$ 
and $\sigma_v= 17.4\kms$, overlaid for reference.
\label{f:histovel}}
\end{center}
\end{figure}

The stars in Table \ref{t:stars} are sorted by increasing projected
geometrical radial distance from the centre of Tucana.  The final
sample of most probable members excludes: the 3 obvious likely
non-member stars with velocities below $150\kms$; star \#10, which
is clearly a double star in the imaging data and also has slit
mis-centering problems; star \#21, which does not have any accurate
photometric information; and the star \#22, which lies outside the
tidal radius.  For the purposes of analysis we exclude the
above 6 stars, but we note that \#10, \#21 and \#22 are still potential
Tucana members. If \#21 and \#22 are included in the analysis
there is no significant change to any of the results.

%Figure 5
\begin{figure}
\begin{center}
\includegraphics[width=5.5cm, angle=-90]{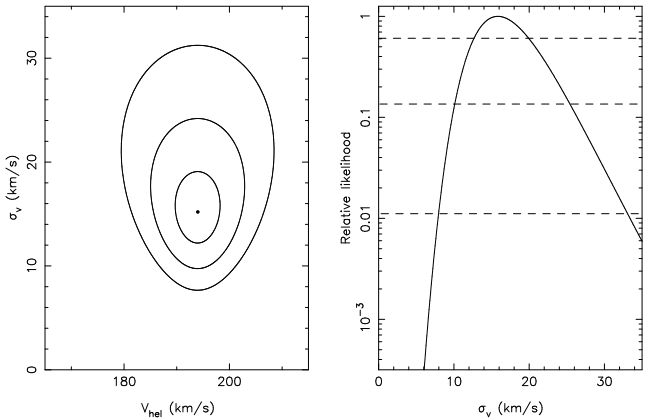}
\caption{
Left panel: the 2D maximum likelihood solution for the 17 members of
Tucana lying within a projected geometric radius of 125~arcsec, 
contours are at the 1, 2 \& 3 $\sigma$
levels.  The right panel shows the marginalised (with respect to
systemic velocity) ``central'' velocity dispersion solution, with
dashed lines denoting the 1, 2 \& 3 $\sigma$ confidence levels.
\label{f:mlvel}}
\end{center}
\end{figure}

Fig.\ \ref{f:histovel} shows the distribution with projected geometric
radius (elliptical radius) for all 23 stars observed (left hand
panel), and a histogram with a Gaussian model fit determined from a
maximum likelihood analysis of the distribution of the radial
velocities of the 17 likely members (right hand panel).  The 2D
maximum likelihood fit together with the marginalised 1D velocity
dispersion is shown in Fig.\ \ref{f:mlvel}.  We find the systemic
heliocentric velocity to be $v_{\rm helio}=+193.0\pm4.9\kms$
(V$_{\rm GC}=+97.9\kms$) with a dispersion of $\sigma=17.4^{+4.5}_{-3.5}\kms$, 
where the error ranges corresponding to 68\%
confidence intervals are derived from the relevant marginalised 1D
distributions.

A straightforward statistical average of the same distribution of
stars gives results very similar to the maximum likelihood fit since
the velocity errors are generally much smaller than the apparent
dispersion.  We also note the apparent non-Gaussian nature of the
distribution compared to the fit,

\subsection{Internal kinematics}
\label{s:kinematics}

\begin{figure}[!ht]
\begin{center}
\includegraphics[width=5.5cm, angle=-90]{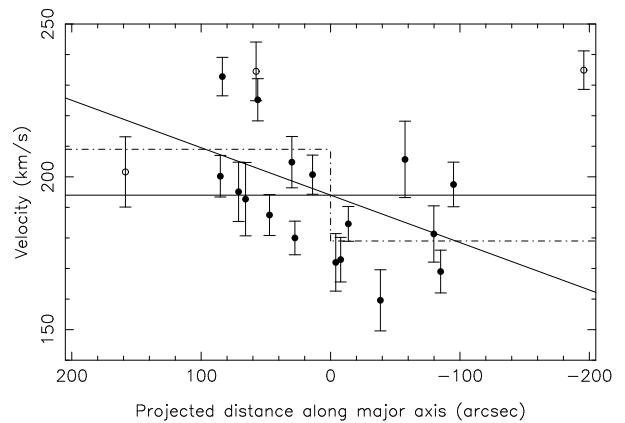}
\caption{
Heliocentric radial velocities of the 20 likely member stars in Tucana
as a function of their projected major axis distance (East is to the
left).  The three open circles are the likely binary star (\#10), the
star lacking photometry (\#21), and the star beyond the tidal
radius (\#22).  The solid line shows a solid body rotation
curve with slope $6.5 \times R/R_{\rm core}\pm2.9 \kms$ fitted to
the filled circle points.  The dot-dashed line shows a fiducial flat
rotation curve with a rotation velocity of $15.0\kms$.
\label{f:rotcur}
}
\end{center}
\end{figure}

Tidal influences are unlikely to be significant at this isolated
location. We examined the velocity distribution as a function of
spatial position, and in particular searched for signs of possible
rotational signatures.  Tucana has a large ellipticity $e=0.48$ with a
position angle running almost parallel to the R.A.\ axis \citep[P.A.\
97~deg,][]{mat98}.  The spatial distribution of spectroscopic
measurements provides reasonable coverage along this major axis but is
somewhat more restricted along the minor axis (see Figure
\ref{f:stars}).  With this caveat, the most prominent correlation
found was with respect to projected major axis position. Fig.\
\ref{f:rotcur} illustrates this and shows the radial velocities of
probable Tucana member stars as a function of their projected major
axis distance.  The dot-dash line shows a fiducial flat rotation curve
of $15.0\kms$ in the observed frame, and the solid line the fitted
slope, $6.5\times R/R_{\rm core}\pm2.9\kms$, of a solid body
rotation curve constrained to pass through the systemic velocity at
the origin.

The reliabilty of our measurements and the accompanying errors
were vigourously tested. This included independently analysing data
taken for Tucana on different nights to compare the results. We also
tested that our method did not find a rotation signal due to the
alignment of the CCDs on the sky, or our data analysis techniques by
analysing data from the FORS2 archive for observations of fields in
the Large Magellanic Cloud and in a number of globular clusters. 

These apparent rotation signatures we find in Tucana are comparable
with values recently found for other dSphs e.g.\ Cetus \citep{lew07}
and Sculptor \citep{B08a}.

%Figure 6

\subsection{Mass:to:Light ratio}
\label{s:moverl}

The photometric structural parameters of Tucana presented in
\citet{sav96} imply a total luminosity of $5.6\pm1.6\times10^5\lo$.  
Several different estimates of the central surface brightness,
$\Sigma_o$, are given by \citet{sav96} so for convenience we have used
the King profile fit central surface brightness of $24.76\pm0.18$
mag/arcsec$^2$ as a fiducial value in this section.  At a distance of
880 kpc this translates to a central surface brightness of $4.3 \pm
0.8 \lo$~pc$^{-2}$, although we caution that the errors may be
underestimated given the range of possible central surface brightness
values.

Following \citet{rich86} we can estimate the mass-to-light (M/L) ratio
by making the simplifying assumption that mass follows light in this
system and then the total M/L is given by
\begin{equation}
      {M \over L} = \eta {9 \sigma_v^2 \over 2 \pi G \Sigma_o r_h }
\end{equation}
where $\eta$ is a dimensionless constant of order unity for a wide
variety of different light profiles.  The appropriate half-light
radius $r_h$ = 221~pc (or 51$''$) we define using the geometric King 
core, $R_{core}$, and tidal radii, $R_{tidal}$ listed in Table
\ref{t:tucana}. $\sigma_{\rm v}$ is the central velocity
dispersion, not including stars \#10, \#21 \& \#22.  This gives a
mass-to-light ratio of $M/L=105^{+95}_{-49}$, where the errors
include a contribution from uncertainties in both the kinematics and
the photometry.  This implies a total mass of the system of 
$\approx6\times10^7\mo$.

However, as noted in the previous section, there is a systematic trend
in velocity as a function of position along the major axis, possibly
due to inherent rotation.  If we remove the ``rotational'' signature
using a linear fit to the marginalised 1D maximum likelihood solutions
the systemic velocity becomes $v_{\rm helio}=+194.0\pm4.3\kms$
(V$_{\rm GC}=+98.9\kms$) and the velocity dispersion 
$\sigma_v=15.8^{+4.1}_{-3.1}\kms$.  This implies a pressure supported mass of
$\approx5\times10^7\mo$.  The rotational velocity is $\approx16\kms$ 
at a radius of $\approx2\times r_h$, beyond which we have no
statistics (see section~3.1).  However, we also note that for the King
model, 87\% of the light is enclosed within this radius and, if mass
follows light in this system, this is also 87\% of the mass.  The
equivalent rotationally supported mass within $\approx2\times r_h$
is $\approx3\times10^7\mo$, which is comparable to the pressure
supported mass.

If we have detected a genuine signature of rotation, then it implies
that the ratio of rotational support to pressure support in this
system is of order unity. This value lies between what is typical for
classical dSphs and dwarf irregular (dIrr) galaxies.  It is
notoriously hard to measure rotation in systems where it is of a
similar order to the velocity dispersion. This means that not many
dwarf systems have been studied with the required detail.  These
results for Tucana are broadly similar, although somewhat higher,
to what has been found in Sculptor dSph \citep{B08a} and Cetus dSph
\citep{lew07} which are similar luminosity (and mass) systems to
Tucana, with similar star formation histories.  As with Cetus and
Sculptor the ellipticity of Tucana is broadly consistent with being
flattened by rotation \citep{B78}.  Looking at higher luminosity dwarf
irregular galaxies the kinematic studies have been carried out
exclusively with the \hi\ gas. This means that the \hi\ kinematics,
specifically the velocity dispersion, cannot be directly compared to
stellar velocity dispersion measures.  Stellar kinematics are
dominated by the mass of the galaxy and gas kinematics are most likely
dominated by small scale interstellar medium processes, as witnessed
by the fact that basically any galaxy from dIrr to large spirals,
regardless of mass or rotation velocity, appears to have the same \hi\
velocity dispersion $\approx 10 \kms$, \citep[e.g.,][]{mat98, lo93}.

\subsection{Metallicity}
\label{s:metallicity}

We have estimated the metallicities of the observed stars in Tucana
using the well-studied correlation between the equivalent widths (EW)
of \ca\ lines and \fe\ abundance, first outlined by \citet{arm91}.
The validity of this approach over the range $-2.5 \rm <$ \fe $< -0.5$
has been firmly established through the comparison of \ca\ EWs with
high resolution metallicities for a large number of stars in globular
clusters \citep{rut97} and in dSph \citep{B08b}.  We used the
preferred calibration determined by Battaglia et al. according to
which the \fe\ abundance is given by:

\begin{equation}
{\rm [Fe/H]} = -2.81 + 0.44 \left({\rm EW}_{8542} + {\rm EW}_{8662} +
0.64 (V-V_{\rm HB})\right)
\end{equation}

where ${\rm EW}_{\lambda}$ is the equivalent width of the line at
wavelength $\lambda$ (the two brightest \ca\ lines), $V$ and $V_{\rm
HB}$ are the V-band apparent magnitudes of the star and of the
horizontal branch ($V_{\rm HB} = 25.4$) of the system respectively,
see also \citep{cole04}.  In cases where no V-band data were available
we used the conversion V$=({\rm B}+{\rm R})/2$.

%Figure 7
\begin{figure}[!ht]
\begin{center}
\includegraphics[width=6cm, angle=-90]{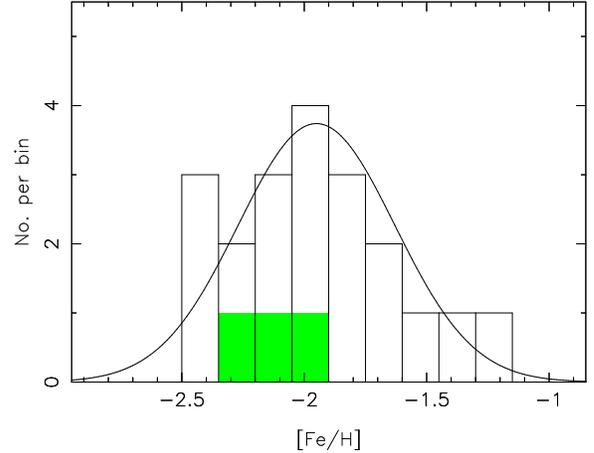}
\caption{
Distribution of metallicities for the 20 likely member stars in Tucana
estimated from \ca\ triplet fitting. The 3 stars which were removed
from the kinematic analysis are shown shaded.  The
solid curve shows an overlaid maximum likelihood fit of a Gaussian
model with a mean metallicity of $\fe = -1.95\pm0.15$~dex and
dispersion (sigma) of $0.32\pm0.06$~dex.
\label{f:histomet}
}
\end{center}
\end{figure}

Fig.\ \ref{f:histomet} shows the histogram distribution of the
metallicities.  The solid curve is a maximum likelihood fit of a
Gaussian function normalised and overlaid on a histogram of all 20
potential Tucana members\footnote{We used a fiducial value of V $=$
22.5 for the star with no photometry since the signal-to-noise in the
spectrum is comparable on average with stars of this magnitude.} . The
likelihood fit indicates a mean metallicity of $\fe = -1.95\pm0.15$
and a dispersion of $\sigma=0.32\pm0.06$~dex.  Similar, although
slightly higher, values for the average metallicity have been found by
\citet{sav96} and \citet{cas96} from comparing the CMD of Tucana with
those of globular clusters.  The relatively large spread in
metallicity that we found is partially due to the errors in the
determination of the equivalent widths and partially to a genuine
intrinsic dispersion in metallicity.  From Table \ref{t:stars} it can
be seen that the typical error in EW is $\approx 0.25$ \AA\, which
translates into an error in metallicity of $\approx$0.1~dex.
Calibration errors of $\approx$0.1~dex tend to affect all stars equally
over this sort of metallicity range \citep{B08b}.  This suggests that
most of the observed metallicity spread is intrinsic to Tucana and
that an extended period of star formation, albeit $\approx$10~Gyrs
ago, is the most likely explanation.

There is no evidence in our spectroscopic sample for a metallicity
gradient in Tucana, as has been seen in the variable stars
\citep[e.g.,][] {bern08}, or in spectroscopic studies of other dSph
galaxies (e.g., Sculptor: \citet{et04}; Fornax: \citet{B06})

%Figure 8
\begin{figure}[!ht]
\begin{center}
\includegraphics[width=0.6\columnwidth, angle=-90]{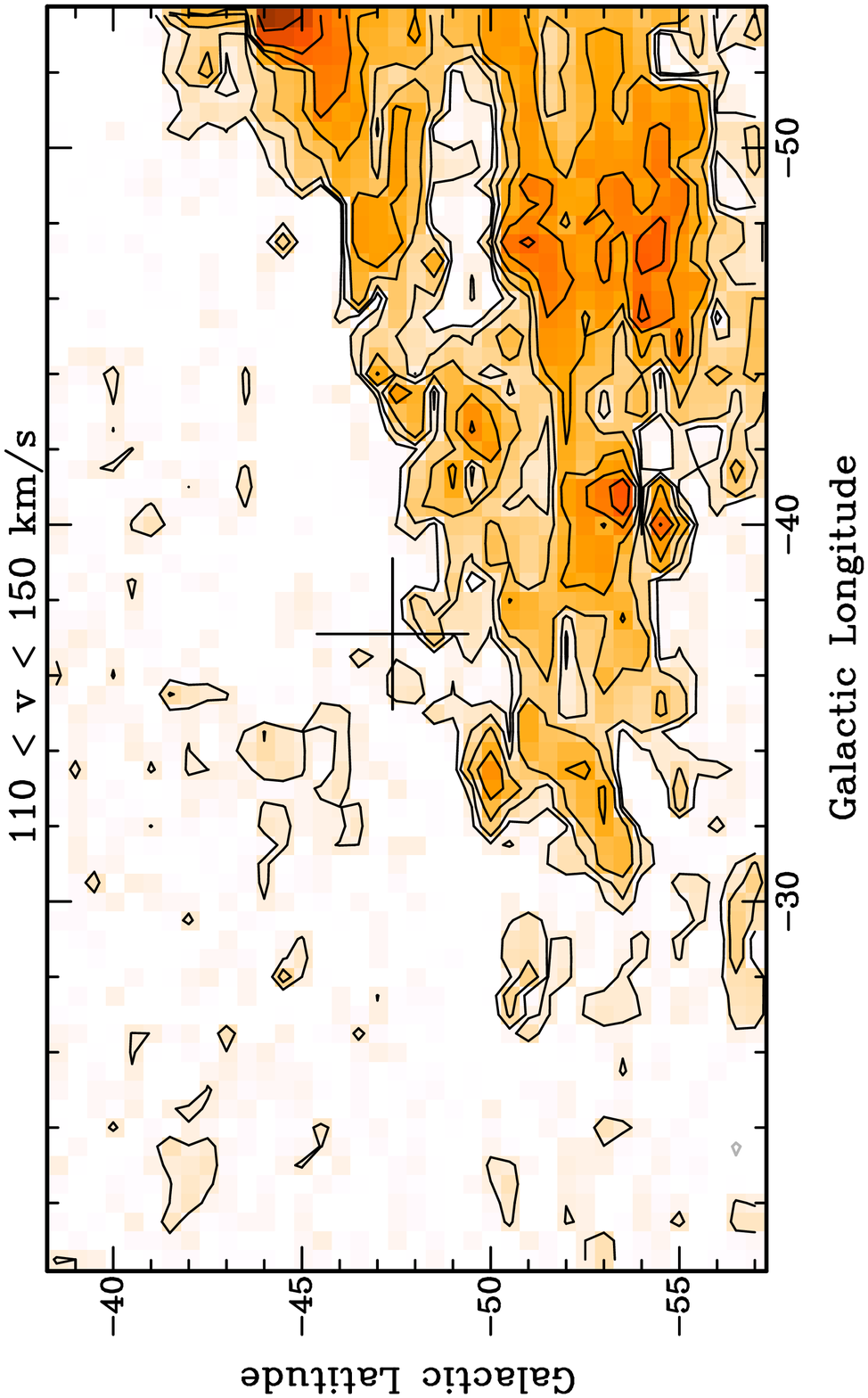}
\includegraphics[width=0.6\columnwidth, angle=-90]{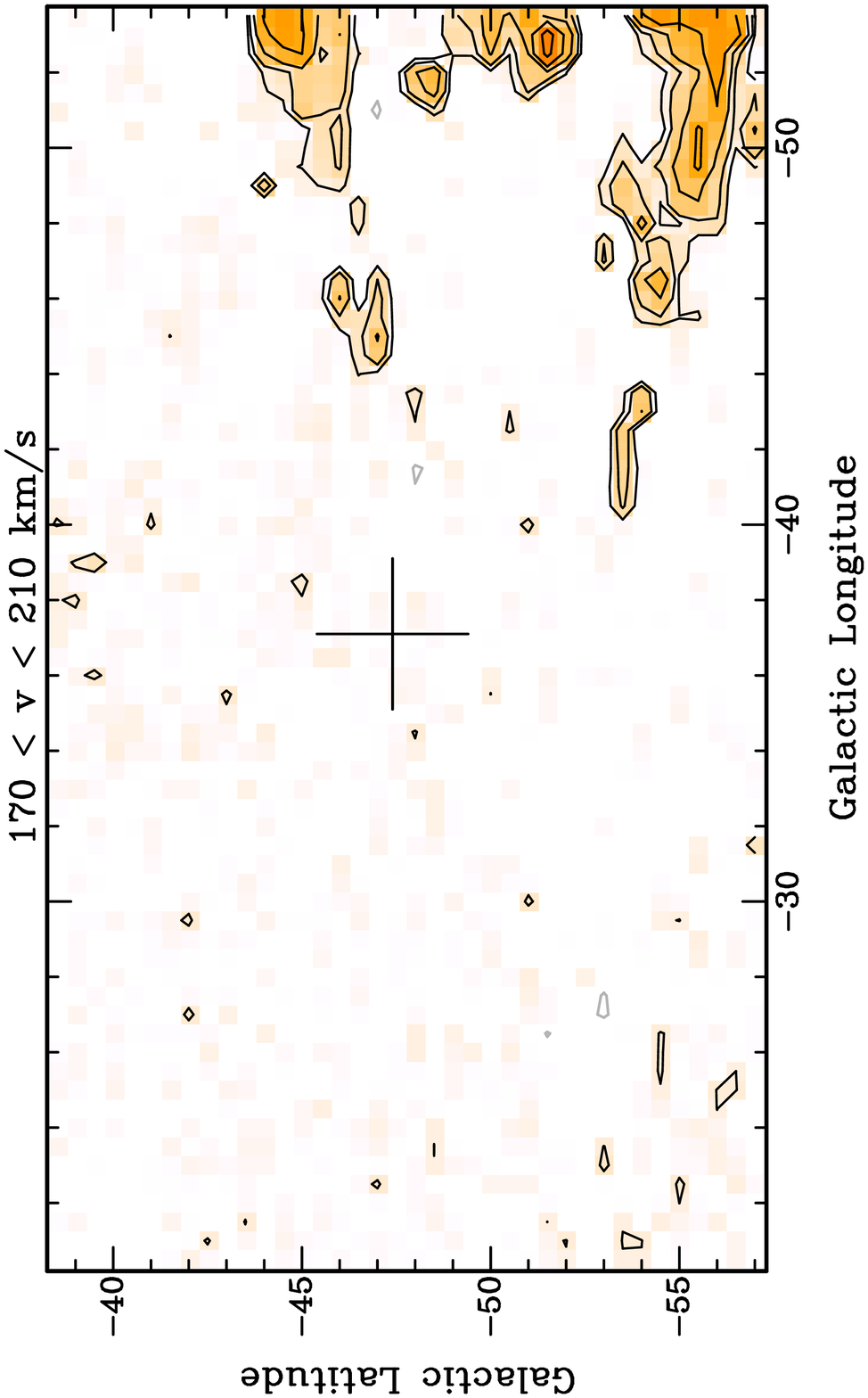}
\caption{
Neutral hydrogen emission in the field around Tucana (cross) from the
LAB \hi\ survey.  The two maps are integrated over $40\kms$ around
two different l.o.s.\ velocities: $130\kms$ (top panel) and 
$190\kms$ (bottom panel).  Most of the emission visible here comes from
the Magellanic Stream; the SMC is at b$=-57.2$, l$=-44.3$, just
outside the frames.  Contours levels: $-2.5$, 2.5, 5, 10, 20, 40, 80,
160 in units of $\sigma$.
\label{f:hi}
}
\end{center}
\end{figure}

\section{Discussion}
\label{s:discussion}

One of the obvious features that distinguishes dSph galaxies from
other dwarf galaxies, such as dIrrs, is the lack of
gas and ongoing star formation. The region of sky around Tucana has
been observed at 21-cm wavelength by \citet{oos96}. They discovered an
\hi\ cloud at a projected distance of 15$'$ to the north-east of
Tucana with a weighted mean heliocentric velocity of $\approx130 \kms$.
Although the cloud has a head-tail shape pointing at Tucana, Oosterloo
et al.  considered the association with Tucana not to be very
probable.  On the other hand, \citet{blitz00} reviewed \hi\ detections
which are possibly associated with dSph galaxies and described Tucana
as a probable \hi\ detection as do \citet{bouch06} using new
observations with the Parkes telescope.

With our determination of the systemic velocity of Tucana we can
firmly exclude an association between \hi\ and Tucana.  From the
channel maps, published by Oosterloo et al., the \hi\ emission of the
cloud has already completely disappeared at a heliocentric velocity of
$165\kms$, which is $30\kms$ away from the velocity of Tucana.  Most
probably, the \hi\ cloud is associated with the Magellanic Stream.  In
Fig.\ \ref{f:hi} we show two \hi\ channel maps, obtained from the
Leiden-Argentina-Bonn (LAB) \hi\ survey of the Milky Way
\citep{kal05}.  The top panel shows an integration of $40\kms$ around
the heliocentric velocity of $130\kms$, which is the velocity of the
Oosterloo et al.'s cloud (the cross shows the position of Tucana).
The cloud (which is not visible in Fig.\ \ref{f:hi}) lies at the edge
the \hi\ emission coming from the Magellanic Stream. At our newly
determined helio-centric optical velocity of Tucana of $\approx
190\kms$ (bottom panel) the foreground emission has totally
disappeared and if there were \hi\ associated with Tucana, it would be
clearly separated from the $local$ emission.  The more sensitive
observations of Oosterloo et al.\ did not detect any emission at these
velocities so we can conclude that Tucana has no neutral hydrogen
associated with it down to their detection limit of 
$1.5\times10^{4}\mo$ (5$\sigma$).

%Figure 9
\begin{figure*}[!ht]
\begin{center}
\includegraphics[width=\columnwidth, angle=-90]{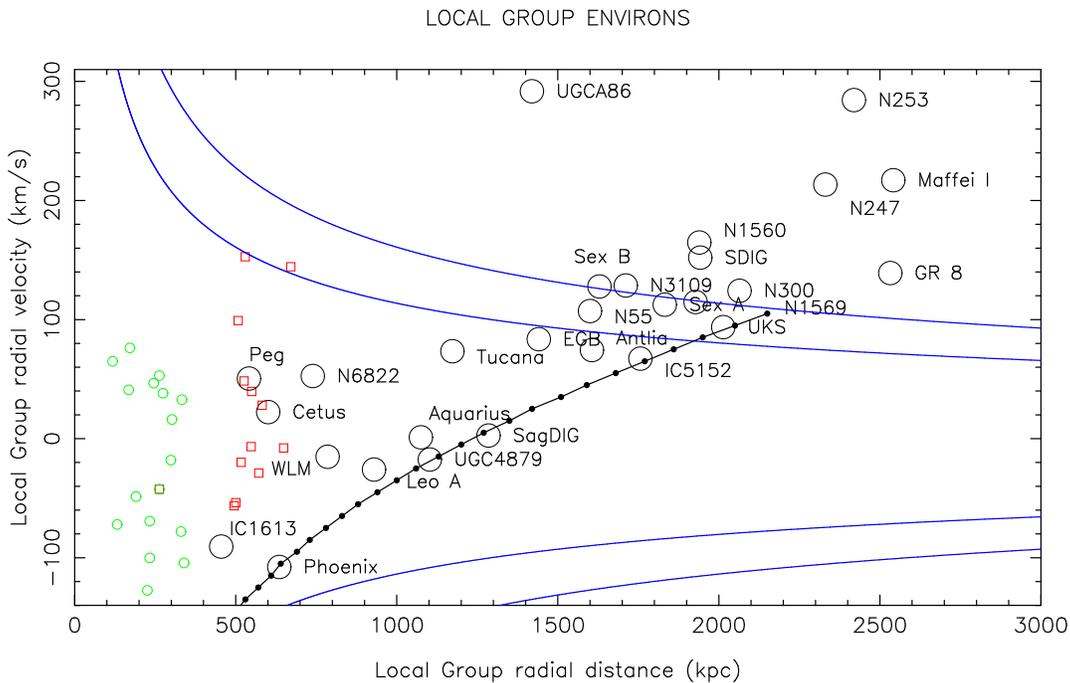}
\caption{
The barycentric Local Group distance versus Local Group radial
velocity for various nearby galaxies in the local universe, including
the new data for Tucana.  The upper and lower tracks are virial
boundaries for purely radial motion and Local Group masses of $3.0
\times 10^{12} \mo$ and $4.5 \times 10^{12} \mo$ respectively.  The
``knotted'' line denotes the limiting distance attainable for an
object with purely radial motion in a notional 13~Gyrs since the Local
Group origin.  Tucana lies close to a bridge of galaxies leading to
the Sculptor group, both in this projection and also spatially in the
Local Group$+$Sculptor Group member galaxy distributions.
\label{f:member}
}
\end{center}
\end{figure*}

The lack of neutral hydrogen is consistent with the lack of recent
star formation in Tucana.  A key question is how did Tucana lose its
gas? The two main possibilities are tidal/ram-pressure interactions
with large galaxies \citep[e.g.,][]{may01} or gas blow-out from
stellar feedback \citep[e.g.,][]{low99}.  The proximity of most of the
dSphs to the large members of the Local Group shows that the former
mechanism must play an important role.  However, given that Tucana is
at a large distance from the Milky Way and even further from M\,31,
one would not expect interaction to play a role in removing its gas.
For example, one of Tucana's closest neighbours is the Phoenix dwarf
galaxy, at a separation of 560~\kpc, but along a similar line of sight
although with a very different radial velocity, at $-13\pm9\kms$
\citep{IT02}.  Phoenix is generally considered to be a transition-type
object, in the process of losing its gas \citep{Young07}.

The heliocentric velocity, $v_{\rm helio} =+194.0 \kms$, of Tucana
corresponds to a velocity with respect to the barycentre of the Local
Group of $v_{\rm LGSR}=+73.3 \kms$ (taking the mass ratio of
M~31:Milky Way as 2:1).  Thus, given the timescales and distance
involved, Tucana has not reached its apocentre (if bound) and it is
still moving away from the Local Group barycentre faster than most
other galaxies at similar distances (e.g., SagDig and Aquarius).  In
Fig.\ \ref{f:member} we show the Local-Group-centric velocity of dwarf
galaxies versus their distance from the Local Group barycentre
\citep[updated from][]{lew07}.  Tucana falls in the region believed to
be associated with the bridge of nearby galaxies toward the Sculptor
Group (on the right side of the plot).  Given that the Sculptor Group
is only 30~degrees away from Tucana, it is quite possible that Tucana
belongs to a bridge (or a cosmological filament?)  connecting the
Sculptor group with the Local Group.

Another possibility arises from interpreting the residual velocity of
Tucana as a receding speed from the Local Group or from the Milky
Way.  In this scheme Tucana may be in a very elongated orbit but still
bound to the Local Group.  If we interpret its velocity as receding
from the Milky Way ($v_{\rm GSR}=+98.9 \kms$), we can in principle
extrapolate back to an encounter with our Galaxy at a $t_{\rm
enc}\lsim 10$~Gyrs ago.  Interestingly, this is consistent with the
epoch of the last burst of star formation in Tucana \citep{sav96}.  A
fly-by Tucana-Milky-Way could have removed the remaining gas from the
dwarf galaxy preventing further star formation.

It has also been shown from simulations that three-body interactions
between dwarf galaxy satellites falling in towards a parent galaxy can
result in the ejection of the smaller dwarf on a nearly radial orbit
at high velocity \citep{sal07}.  These simulations show that a dwarf
galaxy ejected at $300\kms$ from a pericentre approach of about 60 kpc,
is still travelling outwards 7$-$8~Gyr later, at nearly 1~Mpc away from
the primary.  This kind of ejection process might have resulted in the
gas and stars parting ways, and a highly elliptical system. This is
broadly consistent with what we observe for Tucana. A second
effect of this ejection may be the presence of streaming motions
\citep[see also][]{mateo08}.  The velocity gradient observed in
Tucana, can in principle be the signature of streaming instead of
rotation. However, given its position and velocity, Tucana must move
almost radially with respect to the Local Group and the Milky Way
which suggests any tidal distortion, or debris, should be closely
aligned along the line-of-sight.  It therefore seems unlikely for a
tidal disturbance to be visible along the major axis perpendicular to
the likely direction of motion.

\section{Conclusions}

With VLT/FORS2 we have studied the properties of 20 individual
stars which are probable velocity members of the Tucana Dwarf Galaxy,
and we have found that:

i) Tucana has a receding systemic velocity with respect to the Sun
($v_{\rm hel} = +194.0\pm4.3\kms$), our Galaxy ($v_{\rm GSR} = +98.9\kms$
and the Local Group $v_{\rm LGSR} = +73.3\kms$.  Its velocity
dispersion is $17.4\kms$ uncorrected for possible rotation; falling
to $15.4\kms$ after removing a linear rotation signature.

ii) Tucana has no cold gas associated, the upper limit for its gas
content coming from \citet{oos96} is $1.5 \times 10^4 \mo$.

iii) All the stars that we observed have low metallicity with a mean
value of \fe$= -1.95$ and a dispersion of 0.32~dex.

iv) There is evidence for regular rotation around the photometric
minor axis with a velocity up to v$_{\rm rot} \approx16\kms$ at
$\approx2\times r_h$.

In short, Tucana has physical properties similar to the other dwarf
spheroidals in the Local Group but a peculiar location and peculiar
receding velocity.  It may be in a very elongated orbit with respect
to the Local Group or the Milky Way in which case it could have lost
its gas (and stopped forming stars) at its passage through pericentre
close to the Milky Way, $\approx10$~Gyr ago.  Another possibility is that it
resides in a region bridging the Local Group and the Sculptor Group.

\acknowledgements
We thank Gianni Marconi and Thomas Szeifert for their help with the 
observations.

\bibliographystyle{aa}

\end{document}